\documentstyle[preprint,aps]{revtex}
\draft
\begin{document}
\title{\mbox{\hspace*{-0.5cm}On the self--trapping problem 
%\title{\mbox{\hspace*{-0.5cm}On the self--trapping problem 
of electrons or excitons in one dimension}}
\author{G. Wellein and H. Fehske } 
\address{Physikalisches Institut, Universit\"at Bayreuth,
D--95440 Bayreuth, Germany}
\date{Bayreuth, 19 December 1997}
\maketitle
\input{epsf}
\def\gsim{\hbox{$\lower1pt\hbox{$>$}\above-1pt\raise1pt\hbox{$\sim$}$}}
\def\lsim{\hbox{$\lower1pt\hbox{$<$}\above-1pt\raise1pt\hbox{$\sim$}$}}
\def\cH{{\cal{H}}}
\def\cS{{\cal{S}}}
\def\mD{{\mit{\Delta}}}
\def\mP{{\mit{\Psi}}}
\def\ep{\varepsilon_p}
\def\ho{\omega_0}
\def\bk{K}
\begin{abstract}
We present a detailed numerical study of the one--dimensional 
Holstein model with a view to understanding the self--trapping 
process of electrons or excitons in crystals with short--range
particle--lattice interactions. Applying a very efficient 
variational Lanczos method, we are able to analyze the 
ground--state properties of the system in the weak-- and 
strong--coupling, adiabatic and non--adiabatic regimes 
on lattices large enough to eliminate finite--size effects.
In particular, we obtain the complete phase diagram and 
comment on the existence of a critical length for self--trapping 
in spatially restricted one--dimensional systems. In order
to characterize large and small polaron states we calculate  
self--consistently the lattice distortions and the 
particle--phonon correlation functions.  
In the strong--coupling case, two distinct types of 
small polaron states are shown to be possible 
according to the relative importance of static displacement 
field and dynamic polaron effects. Special emphasis is on the 
intermediate coupling regime, which we also study by means of 
direct diagonalization preserving the full dynamics 
and quantum nature of phonons. The crossover from large to small
polarons shows up in a strong decrease of the kinetic energy  
accompanied by a substantial change in the optical absorption spectra.
We show that our numerical results  in all important limiting cases 
reveal an excellent agreement with both analytical perturbation theory 
predictions and very recent density matrix renormalization group data. 
\end{abstract}
%\pacs{PACS number(s): 71.38.+i}
\thispagestyle{empty}
%\newpage
\noindent
\section{Introduction}
Electrons, holes or excitons delocalized in a perfect rigid lattice 
can be ``trapped'' within a potential well produced by displacements of  
atoms from their carrier--free equilibrium positions, provided
the particle--lattice interaction is sufficiently 
strong~\cite{Fi75,Ra82,SW93,Emi95}.  
Trapping of a carrier in this manner is more advantageous energetically 
as compared to wide--band Bloch states, if the lowering of the carrier's 
energy due to its binding exceeds the strain energy required to produce 
the trap. Since the potential that binds the carrier depends on the 
carrier's state itself, i.e., the local distortion of the lattice 
is self--induced by the particle, this process is called 
``self--trapping'' or ``auto--localization''. 
Obviously, self--trapping (ST) is an highly non--linear phenomenon.  
A self--trapped state is referred to as ``large'' if it extends
over multiple lattice sites. Alternatively, if the quasiparticle  
is practically confined to a single site the ST state is
designated as ``small''.  Nonetheless, ST does not imply
a breaking of translational invariance. ST eigenstates 
in a crystal are Bloch--like. Thus coherent transport 
of ST particles can, in principle, occur but the width of the 
corresponding band is extremely small (cf. the discussion 
in~\cite{BEMB92}). 

Introducing the concept of polarons into physics, 
the possibility of electron immobilization or ST was  
pointed out by Landau as early as 1933~\cite{La33}. Shortly after   
the ST of excitons was also suggested and studied 
theoretically~\cite{Pe32,Fr36}.
ST polarons and excitons can be found in a wide variety of alkali 
metal halides, alkaline earth halides, II--IV and group IV semiconductors,
condensed rare gases, organic molecular crystals, electrochromics, 
and other systems~\cite{SW93,SS93}. With the observation of polaronic 
effects in new materials exhibiting exceptional properties
such as the high--$T_c$ cuprates~\cite{SAL95} or the colossal
magnetoresistance manganates~\cite{MLS95}, research on polarons 
has attracted renewed attention. 

Although the problems of exciton and electron ST have much in common
there are fundamental differences. Most notably  excitons are 
short--living non-equilibrium quasiparticles being   
immediately after the optical excitation in the free state 
and can reach the ST state only by tunneling through the potential 
barrier at low temperatures. Moreover, if the electron and hole, forming
the exciton, have very different effective masses, the internal coordinates  
of this large--radius (Wannier--Mott) exciton will be of importance.
 
It is clear that the microscopic structure of the ST state 
is very diverse in various groups of materials.
The stability of different types of ST states depends on the nature 
of the electron/exciton--phonon (EP) coupling 
(e.g. deformation potential~\cite{Ho59a} against Fr\"ohlich~\cite{Fr54} 
coupling), the vibrational frequencies (e.g. acoustic vs. optic),
the dimensionality ($D$) of the lattice and other parameters. 
A detailed classification of ST states and ST criteria 
is presented, e.g., in the excellent review of Rashba~\cite{Ra82}.
In particular, from a scaling treatment of a continuum lattice model 
in the adiabatic limit, it has been shown that  
in {\it multi--dimensional} systems $(D>1)$ with only 
{\it short--range non--polar}  
EP interaction a ST carrier always forms a small 
polaron~\cite{TS74,EH76,Emi95}. If there is an energy barrier 
that separates delocalized and spatial confined states, the 
free and ST states can coexist~\cite{Ra57a,To61}, as evidenced, 
e.g. for solid xenon, by the coexistence of two exciton luminescence 
bands (one narrow and the other broad)~\cite{SW93}.
On the other hand, the presence of a {\it long-range polar} EP coupling 
ensures that at least the formation of large polaron states with moderate
lattice deformation becomes possible. 
The picture is qualitatively changed when turning to {\it low--dimensional}
systems. Recently it was shown that, unlike the continuum limit, the 
formation of a ST state in the 2D case  within a model of local EP coupling
is always accompanied by the formation of an energy barrier attributed to the 
lattice discreteness~\cite{KM93}.  The 1D case is essentially 
different from the 2D and 3D ones. In 1D ST proceeds without overcoming a
barrier. To be more precise, the ST state is the ground state of the 
infinite system at any value of the EP coupling. In the
weak EP coupling regime its radius exceeds several times the lattice spacing,
i.e., a large polaron is formed even under the non--polar interaction 
condition~\cite{Ra82}. 

Although the basic concepts underlying the ST transition 
are long standing and the gross features of 
``large'' and ``small'' polarons and excitons 
have been extensively studied, our understanding of the 
ST problem is still incomplete. In particular the physically most important 
crossover regime, characterized by {\it intermediate} EP coupling
strengths and phonon frequencies, is difficult to handle 
theoretically due to the failure of the standard phase transition 
concept~\cite{GL87a,Loe88}. As yet, there exist no 
well--controlled analytical techniques to describe the transition region. 
Other problems, for example, concern the existence of a critical 
length for ST in spatially restricted 1D systems~\cite{Ra94a}, 
the behaviour of the polaron kinetic energy~\cite{MR97}, 
or the spectrum of light excitons
under ST conditions~\cite{Ra82}.  

With these motivations, in this paper we want to 
discuss the ST problem using numerical methods.
The focus is on the Holstein model in one-dimension.
By exact diagonalization of finite systems we analyze various 
ground state and spectral properties of the model. 
Since for the Holstein model exact results are very 
rare~\cite{Loe88,FL95,CPF95} and previous numerical studies
have been limited either to small systems or to a particular parameter 
regime~\cite{RT92,Ma93,AKR94,FRWM95,CSG97} this is a challenge by itself. 
Besides, we hope to gain more insight into the physics of small/large 
polarons and into the nature of the localization--delocalization transition.

The paper is organized as follows: 
In the next section we briefly introduce the Holstein model and 
outline our variational Lanczos approach that allows to study 
the ground state properties for all regimes of parameters 
on large lattices in a very efficient way. 
The numerical results will be presented in Sec. III. More precisely, 
the phase diagram of the Holstein model (A), the  
electron lattice correlations (B), the optical response (C) 
and the kinetic energy (D) will be discussed.  
The principal results are summarized in Sec. IV.
\section{Model and methods}
With our focus on the self--trapping phenomenon in systems 
with only short--range non-polar electron-- or exciton--lattice 
interaction, it is appropriate to consider a model, 
where a single excess carrier is placed in an one--dimensional 
periodic array of identical molecular units, each having an 
internal vibrational degree of freedom.
Introducing electron (exciton) $a_i^{[\dagger]}$ and phonon  
$b_i^{[\dagger]}$ destruction [creation] operators we can write 
Holsteins's molecular crystal model~\cite{Ho59a} in lattice site
representation as 
\begin{equation}
\cH=\hbar\ho\sum_i  ( b_i^{\dagger} b_i^{}+ \mbox{\small $\frac{1}{2}$})
- \sqrt{\ep\hbar\ho}\sum_i
( b_i^{\dagger}  + b_i^{})\,  n_i^{}
-t\sum_{\langle ij\rangle} ( a_i^{\dagger} a_j^{} + a_j^{\dagger} a_i^{}) 
\label{eq-mo}
\end{equation} 
($n_i^{}=a_i^{\dagger} a_i^{}$; below $\hbar=1$). 
In the case of electrons the Holstein Hamiltonian~(1) has been 
studied extensively as a paradigmatic model for small polaron formation. 
Here $t$ denotes the nearest-neighbour free--electron transfer amplitude,
$\ep$ is the strong--coupling polaron binding energy
(in the atomic limit $t=0$), and $\ho$ is the bare phonon 
frequency of the dispersionsless optical phonon mode. 
Referring to excitons we have in mind the small--radius 
(Frenkel or charge transfer) excitons only, and neglect, 
in the lowest order of approximation, the internal structure  
of the exciton, i.e., we consider it as a single neutral particle.   

Generally speaking, the ground--state and spectral properties of 
the model~(1) are governed by three ratios
(control parameters) defined from the bare energy scales 
$t$, $\ho$, and $\ep$. 
First the adiabaticity parameter
\begin{equation}
\alpha =\ho/t
\end{equation}
determines which of the two subsystems, excitons/electrons or phonons,
is the fast and which is the slow one. In the adiabatic limit $\alpha\ll 1$, 
the motion of the particle is affected by quasi--static lattice
deformations (adiabatic potential surface) depending on 
the strength of the particle--phonon interaction. 
On the contrary, in the anti--adiabatic 
limit $\alpha\gg 1$, the lattice deformation is presumed to
adjust instantaneously to the position of the carrier.    
Conveniently the particle is referred to as a ``light'' or ``heavy'' 
electron/exciton in the adiabatic or non-adiabatic regimes~\cite{Ra82}.
The second parameter is the dimensionless EP coupling constant  
\begin{equation}
\lambda =\ep/W\,,
\end{equation}
where  $W=2Dt$ denotes the half--width of the electron (exciton)
band in a rigid $D$--dimensional lattice. Let us stress 
that $\lambda$ represents the ratio between ``localization'' 
energy $(\propto \ep)$ and {\it bare} kinetic energy $(W)$ 
of a single particle. 
Both, $\lambda$ and $\alpha$, are commonly used as parameters 
within a perturbative analysis of the Holstein model in the limits
of weak $(\lambda \ll 1)$ and strong $(\lambda \gg 1)$ EP couplings. 
In the latter case two different approaches, based on 
expansions in powers of $(\alpha \ll 1)$ and $(1/\lambda \ll 1)$, 
have been elaborated for the adiabatic~\cite{Ho59a,Ra57a} and 
non--adiabatic~\cite{Go82} regimes, respectively. 
A third parameter,  
\begin{equation}
g^2=\ep/\ho\,,
\end{equation}
will show to be crucial in the strong--coupling situation.
$g^2$ determines the relative deformation of the lattice 
which surrounds the particle.   
  
In the limit of small particle density,
a crossover between essentially delocalized carriers
and quasi-localized particles is known  to occur
from early quantum Monte Carlo calculations~\cite{RL82}, 
provided that the {\it two} conditions $\lambda \stackrel{>}{\sim} 1$ 
and $g \stackrel{>}{\sim} 1$ 
are fulfilled. So, while the first condition is more restrictive
if $\alpha$ is small, i.e. in the adiabatic case, the formation of 
a small ST state will be determined by the second criterion in 
the anti--adiabatic regime~\cite{CSG97,WF97,FLW97}.

It is not surprising, that the standard perturbative techniques 
are less able to describe the system close to the crossover region,
where the energy scales are not well separated ($\lambda \sim 1,\;
g \sim 1$). Therefore we will apply in the following two distinct  
numerical methods that allow investigating the ST phenomenon
on finite clusters with great accuracy.  

The {\it first method} is a variational Lanczos technique   
(developed originally for the Holstein t--J model~\cite{FRWM95,Fe96}), 
which enables us to study the {\it ground--state properties} of
fairly large systems. In recent work  
this technique has been adapted successfully to treat the lattice
degrees of freedom in the generalized double--exchange Hamiltonian  
commonly used for the description of colossal magnetoresistance 
materials~\cite{RZB96}.
In the case of the pure Holstein model, as a first step, we perform
an inhomogeneous variational Lang--Firsov (IVLF) transformation
\begin{equation}
\label{eq-ut1}
   \tilde{\cH} = {\cal{U}}^\dagger\, \cH \, {\cal{U}},\;\;
   {\cal{U}}= \mbox{e}^{-\cS_1(\mD_i)}\,  
                         \mbox{e}^{-\cS_2(\gamma)}\,  
                         \mbox{e}^{-\cS_3(\tau)}   
\end{equation}
with 
\begin{eqnarray}
        \label{s1}  
\cS_{1}(\mD_i)&=& - \frac{1}{2 g\alpha} 
               \sum_i  \mD_i
               ( b_i^{\dagger} -  b_i^{} )\,,\\[0.2cm]               
\label{s2} 
\cS_{2}(\gamma)&=&-\gamma g 
                \sum_{i} 
  (b_i^{\dagger} -  b_i^{})\, a_i^{\dagger}a_i^{}\,,\\[0.2cm] 
\label{s3}                              
   \cS_{3}(\tau)&=&- \ln \tau^{-1/2} \sum_i (b_i^{\dagger} b_i^{\dagger} 
                       - b_i^{} b_i^{} ) \,,                  
\end{eqnarray} 
(rescaling $\cH=\cH/t$ and measuring, in what follows,  
all energies in units of $t$). In a certain sense the canonical transformation 
${\cal U}$ is a (variational) synthesis of two different approaches 
developed in the adiabatic~\cite{La33,Pe46a,Ho59a,Ra57a} 
and anti--adiabatic~\cite{Pe32,Fr36,LF62} theories of ST
polarons/excitons. $\cS_{1}$ introduces a set of static 
site--dependent displacement fields $\mD_i$
related to local lattice distortions. This transformation 
ensures the correct behaviour in the adiabatic limit. 
That is, $\cS_{1}$ describes the ST of ``light'' 
excitons/electrons under the conditions that (i) the 
electronic bandwidth significantly exceeds
the characteristic phonon frequency and (ii) the lattice deformation
energy is large (which allows one to treat the lattice vibrations
quasi--classically). Within polaron theory such type of ST quasiparticle is 
often called adiabatic Holstein polaron (AHP)~\cite{Ho59a}.  $\cS_2$, 
on the other hand, describes the ST process
in the anti--adiabatic limit. The variational parameter $\gamma$  
(with $0\leq \gamma \leq 1$) is introduced as a measure of the 
non--adiabatic phonon dressing of ``heavy'' particles, 
designated as ``localized'' excitons~\cite{Da62} or non--adiabatic 
Lang--Firsov small polarons (NLFP). For $\gamma =1$, the well--known
Lang--Firsov displaced--oscillator transformation results. 
In addition, we have applied the two--phonon squeezing transformation
$\cS_3$ ($0\leq \tau \leq 1$)~\cite{Zh88a,FCP90}. The squeezing phenomenon
is a many particle effect being of special importance at intermediate EP 
coupling strengths.  This effect can be seen as a phonon frequency softening
and tends to offset the (polaron) band narrowing.
As a second step, we approximate the eigenstates $|\tilde{\mP} \rangle$ of 
$\tilde{\cH}$ by the variational states  
\begin{equation}
|\tilde{\mP}_V \rangle =  
|\tilde{\mP}_{ph} \rangle \otimes|\tilde{\mP}_{el} \rangle\,.
\label{eq-va2}
\end{equation} 
Then, performing in a third step the average over the transformed 
phonon vacuum,  
$\bar{\cH}\equiv \langle\tilde{\mP}_{ph}^0 |\tilde{\cH} 
|\tilde{\mit \Psi}_{ph}^0 \rangle$, we obtain the effective
(electronic/excitonic) Hamiltonian 
\begin{eqnarray}
\label{heff}
 \bar{\cH}  &=&  \frac{\alpha}{4}
(\tau^2+\tau^{-2})N\,+\,\frac{1}{8\lambda}
\sum_i \mD_i^2 \,-\, \left(1-\gamma\right)
         \sum_{i} \mD_i^{} n_i^{} \nonumber\\[0.2cm]
& &   
- 2\lambda\gamma(2-\gamma) \sum_{i} n_i^{}   
        \,-\,\mbox{e}^{-g^2\gamma^2\tau^2}\sum_{\langle ij\rangle} 
( a_i^{\dagger} a_j^{} + a_j^{\dagger} a_i^{}) \,.
\end{eqnarray}
In (10), the first term leads to an increase of the zero--point energy 
of the phonons if $\tau^2< 1$. 
The second and third contributions give the elastic energy and 
the particle--lattice interaction, respectively, both owing to the static
lattice deformation. As a result of the incomplete LF transformation
we get a constant (polaronic) level shift and an  
exponential band renormalization (fourth and fifth terms).
Even the simplified model (10) cannot be solved exactly.
Therefore we carry out a Lanczos diagonalization on finite $N$--site
lattices using periodic boundary conditions. 
Employing the Hellmann--Feynman theorem, the $N+2$ variational parameters 
can be obtained by iteratively solving the following set of 
self--consistency equations: 
\begin{eqnarray}
\mD_i&=&4\lambda (1-\gamma)\,\bar{n}_i\,,\\[0.2cm]
\gamma&=&\frac{\alpha [ 1 - \bar{E}_{\mD n}/4\lambda] }{\alpha
- \tau^2 \mbox{e}^{-g^2\gamma^2\tau^2} \bar{E}_{kin}}\,,\\[0.2cm]
\tau^2&=&\frac{\alpha}{\sqrt{\alpha^2-8\lambda\gamma^2
\mbox{e}^{-g^2\gamma^2\tau^2}\bar{E}_{kin}/N}}\,,
\end{eqnarray}
where
\begin{eqnarray}
\bar{n}_i\;\;&=&  \langle
n_i\rangle_{\bar{\cH}}^{}\;\;\;\mbox{with}\;\;\; \sum_i\bar{n}_i=1 \,,\\
\bar{E}_{kin}\,&=& -\sum_{\langle i j \rangle} 
\langle (a_{i}^\dagger a_{j}^{} + a_{j}^\dagger a_{i}^{} ) 
\rangle_{\bar{\cH}}^{}\,,\\
\bar{E}_{\mD n}&=&\sum_{i} \langle \mD_i n_i \rangle_{\bar{\cH}}^{}\,, 
\end{eqnarray}
denote the local particle density, the kinetic energy, 
and the EP interaction contribution to the ground state energy, respectively.
Note that each iteration step involves the exact diagonalization of 
$\bar{\cH}(\gamma,\tau^2,\{\mD_i\})$. 
We note further that the Hamiltonian (10) potentially contains symmetry--broken
states which originate from inhomogeneous displacement fields $\mD_i\neq 0$.
Therefore we have to work with an unsymmetrized set of basis states.

The {\it second method} we are going to use in the computational work  
is the direct numerical diagonalization  of the initial 
Holstein Hamiltonian~(1). On the one hand this should bring
out valuable information on the applicability of various approximative
analytical and numerical approaches.  In particular we would like to test 
the quality of IVLF--Lanczos scheme described so far. On the other hand,
combining our exact diagonalization (ED) algorithm 
with the Chebyshev recursion and maximum entropy
methods~\cite{Siea96}, we are able to discuss   
{\it dynamical properties} of the systems, e.g. the optical conductivity, 
in more detail. Moreover, ED provides the only reliable tool 
for treating the transition regime. Differently from the IVLF--Lanczos 
treatment the translational invariance of the system is ensured. 
The ED method, based on a well--controlled truncation procedure of the 
phonon Hilbert space, has been described elsewhere~\cite{WRF96,WF97,FLW97}.
Using parallel computers, we are able to diagonalize systems 
with a total dimension of $10^{10}$.
\section{Numerical results and discussion}
\subsection{Phase diagram}
In the numerical work we start with a discussion of the ground--state 
properties of the transformed Hamiltonian~(10). Applying the IVLF--Lanczos 
technique presented in the previous section, we have determined the 
phase diagram of the 1D Holstein model. The results are depicted in Fig.~1. 
First let us consider the regimes II and III, corresponding to
large and small polarons, respectively. Just for brevity we will 
use in the following the ``polaron terminology'', keeping in mind
that all statements hold for the case of Frenkel excitons as well.   
The distinct types of polarons, found  in II and III, may be characterized
by the spatial extension and strength of the (inhomogeneous) 
lattice displacements $\mD_i$ and by the magnitude of the polaron variational 
parameter $\gamma$ (see below). 
From {\it exact} analytical~\cite{Loe88} 
and numerical~\cite{RL82,WRF96,CSG97} (cf. also Sec.~III~D) results 
it is well known that the large--size
polaron turns {\it continuously} into a small--size polaron with 
increasing EP coupling. Since there is no true phase transition
between large and small ST states at $\alpha>0$, the transition line (stars)
shown in Fig.~1 only indicates the crossover region, which gets wider as the 
phonon frequency ($\alpha$) increases. 
Within our IVLF--Lanczos treatment, the transition line has been fixed  
by the criterion $\mD_1/\mD_0=1/e$. Performing a finite--size analysis 
of the II$\rightleftharpoons$III transition, 
we found that the results obtained for the 64--site lattice almost agree 
with the extrapolated values for the infinite system. 
Via Eq. (11), $\mD_i$ is directly related to the polaron density at site $i$.
Of course, this condition can only give a crude estimate of 
the ``transition'' from large to small polarons, 
in particular in the non--adiabatic regime 
where the static $\mD_i$ are less significant. 
According to the importance of the  $\mD_i$ ($\alpha\leq1$) and 
$\gamma$  ($\alpha\geq1$) effects, the small polaron formed in region~III 
will be called adiabatic Holstein polaron (AHP) 
and non--adiabatic Lang--Firsov polarons (NLFP), respectively. 

As a peculiarity of our {\it finite} system a further strongly finite--size
dependent transition to a nearly free electron state~(I) is observed 
by lowering the EP coupling strength. In other words, it seems 
that a critical coupling $\lambda_c(N)$ or equivalently a critical 
system size $N_c(\lambda)$ exists for self--trapping in 1D.  
Indeed, for the 1D {\it continuum model}, where the ST problem 
can be described by a non-linear 
Schr\"odinger equation, it has been shown recently 
by Rashba~\cite{Ra94a} that the ST condition is
\begin{equation}
\lambda > \lambda_c=\pi^2/2 N\,.
\end{equation}
This relation holds rigorously within the adiabatic theory~\cite{Ra94a}
and is reproduced by our {\it lattice model} calculation as well 
(cf. inset Fig.~1). 
At $\alpha =0$, the nearly free electron phase~(I) corresponds to a solution
with  $\gamma=0$ and $\mD_i=\mD=4\lambda/N$. 
The kinetic energy, however, is unrenormalized.
Our IVLF-Lanczos scheme allows to extend the above considerations
to the finite phonon frequency regime. Again, at low EP couplings, 
we found a nearly free electron phase with a small uniform 
level shift ($\propto -2\lambda[\gamma(2-\gamma)+2(1-\gamma)^2/N]$). 
More significantly, since we have $\gamma >0$ now, 
the inclusion of non--adiabatic phonon effects slightly renormalizes 
the electron bandwidth $\bar{W}=4\exp\{-g^2\gamma^2\tau^2\}$.
If $\lambda$ becomes larger than  $\lambda_c(\alpha)$ 
the ST proceeds by a monotonic lowering of the total energy
without overcoming a ST barrier. The scaling of $\lambda_c(\alpha)$ 
with $N$ is shown in the inset of Fig.~1. 
In the thermodynamic limit $N\to \infty$ we obtain 
$\lambda_c(\alpha)\to 0$, i.e., in an {\it infinite} 1D system ST 
takes place at any finite value of the EP coupling.

To elucidate in more detail the different nature of polaronic states 
occurring in the ground--state phase diagram of the effective model~(10),
we present in Fig.~2 the behaviour of the variational parameters. 

First of all, we should emphasize that, 
studying the {\it single--electron} problem, 
the squeezing effect ($\tau^2<1$) is only of minor importance.  
This is obvious from~(13), which yields $\tau^2=1$ 
in the thermodynamic limit.
For finite systems, the leading ($1/N$)--corrections 
$(\propto g^2 \gamma^2 \mbox{e}^{-g^2\gamma^2\tau^2} \bar{E}_{kin}/\alpha N)$  
tends to zero in the weak-- and strong--coupling adiabatic and 
anti--adiabatic limits. 

The spatial extension of the static lattice deformation $(\mD_i)$
is visualized in Figs.~2 (a) and (b) for different
EP couplings corresponding to the nearly free, large and small polaron cases.
As discussed above, the $\mD_i$ are being constant for 
$\lambda<\lambda_c(\alpha)$ (phase~I). For the large--size polaron 
(phase~II), the lattice displacements fits extremely well to the
relation   
\begin{equation}
\mD_i=\mD_0 \,\mbox{sech}^2 [\lambda_{eff}\cdot i\,]\,,
\end{equation} 
where  $\lim_{\alpha\to 0}\lambda_{eff}(\lambda,\alpha) =\lambda$.
It is worth emphasizing, that the functional form~(18), which 
has been derived in the framework of an adiabatic 
continuum theory~\cite{Ra57a,CC92,KM93}, also describes the 
displacement fields in the non--adiabatic  large polaron regime. Obviously, 
$\lambda_{eff}$ defines a characteristic inverse length scale
in the system, i.e., the  radius of the large polaron is approximately 
given by  $\sim (2\lambda_{eff})^{-1}$.  
For $\alpha=0.1$ and $\lambda=0.25$  we got   
$\lambda_{eff}/\lambda\simeq 0.935$. 
On the other hand, at $\alpha=3$ and $\lambda=2.5$, 
the effective coupling becomes
strongly reduced: $\lambda_{eff}/\lambda\simeq 0.116$.
In the strong--coupling regime, we observe an exponential
decay, $\mD_i\simeq\mD_0\mbox{e}^{-i/\xi}$, of the lattice 
distortion away from the polaron site (see insets), where 
$\xi$ denotes the small polaron radius. We found $\xi\simeq 0.29$
(0.19) for $\lambda=1.5$ (5.0) and $\alpha=0.1$ (3.0), i.e., in both cases
the ST state is mainly confined to a single lattice site. 
While, in the framework of our interpolating theory, 
the $\mD_i$ can be taken as a measure of the ``adiabatic character'' 
of the polaronic quasiparticle, its ``non--adiabatic part'' is described 
by the Lang--Firsov variational parameter $\gamma$ shown in Fig.~2~(c).
Of course, in the case of ``light'' electrons $(\alpha < 1)$, the 
non--adiabatic polaron effect is rather small. In particular for $\lambda>1$, 
when the small AHP is formed, $\gamma$ becomes strongly suppressed. 
Here the renormalization of the polaron band is mainly driven
by the static displacement fields $\mD_i$. 
Otherwise, for ``heavy'' electrons, we observe larger values of 
$\gamma$, which increase with increasing EP coupling strength. 
As a result the free--electron band 
is transformed into a renormalized polaron band.
Due to the (generalized) Franck--Condon factor 
$\mbox{e}^{-g^2\gamma^2\tau^2}$, the bandwidth is exponentially small 
under strong coupling conditions ($g^2,\,\lambda \gg1$), where also 
the $K$--dependent corrections to the band dispersion become 
negligible~\cite{FLW97}. Within our variational approach 
we found a first--order transition to the AHP state 
at extremely strong EP interaction ($\lambda_{II/III}\simeq 4$, 
for $\alpha=3$). 
However, it should be noted that this sharp transition 
is in some sense an artifact of our IVLF--scheme that 
compares the ground--state energies of the AHP and NLFP states, 
both obtained in the lowest order of approximation 
(remind that by deriving (10) we have performed the average 
over the {\it zero--phonon} state only). 
Including higher--order corrections,  the NLFP with $\gamma\to 1$  
is stabilized in the non--adiabatic strong--coupling regime 
(cf. the discussion in Sec.~III~D).
\subsection{Electron lattice correlations}
In looking for a characterization of the different polaronic regimes 
for the quantum--phonon Holstein model~(1), 
we have calculated  the (normalized) correlation function between 
the electron position $i=0$ and the oscillator displacement at site $j$,  
\begin{equation}   
\chi_{0,j}=\frac{\langle n_0 (b_{0+j}^{\dagger}
+b_{0+j}^{})\rangle}{2g \langle n_0\rangle}\,, 
\end{equation}
by means of direct diagonalization technique. 
Here fermion and boson degrees of freedom are related by the well--known
relation $\langle b_i^{\dagger}+ b_i^{} \rangle_{\cH}=2g
\langle n_i\rangle_{\cH}$ ($=2g/N$ for the single electron case). 

Alternatively, working with the effective Hamiltonian~(10), 
the electron--lattice correlation function~(19) can be expressed as 
\begin{equation}
\bar{\chi}_{0,j}=\gamma \delta_{0,j} + \frac{\mD_j}{4\lambda}\,.
\end{equation}
Hence we can use the static correlation functions $\chi_{0,j}$ and
$\bar{\chi}_{0,j}$ to test the accuracy of the IVLF--scheme.

Figure~3 shows the static correlation functions~(19) and~(20)
in the adiabatic (a) and non--adiabatic (b) regimes 
for several coupling parameters $\lambda$ corresponding to the
different polaronic ``phases'' indicated in~Fig.~1. 
For parameters close to the adiabatic weak--coupling regime (phase I), 
the amplitude of $\chi_{0,j}$ is extremely small and the 
spatial extent of the electron--induced lattice deformation
is spread over the whole lattice. In the quantum--phonon model~(1) the 
coupling gives rise to a weak dressing of the  electron at any 
finite $\alpha$.  However, the carrier is not trapped 
due to the zero--point quantum lattice fluctuations~\cite{JW97}. 
From~(20) it is clear that in the effective model~(10)
the on--site dynamical polaron and spatially extended 
static displacement field contributions are well separated. 
Neglecting the residual polaron--phonon interaction,  
the IVLF--approach describes the real situation 
by a (variational) superposition of both effects. 
In the adiabatic large polaron region~II a much better description 
of the exact behaviour is obtained. Especially when the small 
AHP state evolves at $\lambda\sim 1$ the IVLF results are even 
in quantitative agreement with the DMRG (density matrix renormalization group)
data obtained very recently by Jeckelmann and White~\cite{JW97} 
(see inset Fig.~3~(a)). 
At this point we would like to emphasize that performing such DMRG--  
and, in particular, ED--calculations requires much more memory and
CPU--time resources than our extreme simple and very fast 
IVLF--computations. In Fig.~3, the system sizes treated within 
the IVLF--scheme have been restricted in order to make possible   
a direct comparison with the available ED/DMRG data. 
Obviously, for intermediate to strong EP couplings,  
the IVLF--results agree almost exactly with the ED and DMRG data
(Fig.~3~(b)). Here the electron--lattice correlations become very 
localized and finite--size effects are less important. 
Although the behaviour of $\chi_{0,j}$, shown in the main part of Fig.~3, 
is found to be very similar for $\lambda=1.5$, $\alpha=3.0$ and 
$\lambda=4.5$, $\alpha=1.0$, we would like to  emphasize that  
both parameter sets describe completely different physical situations.  
The distinct nature of the corresponding small polaron states 
becomes apparent from the variation of the static 
displacement fields shown in the insets.  
For $\lambda=4.5$ and an intermediate (or low) phonon frequency,
we observe a static lattice distortion in the vicinity of the 
electron only. Since $\mD_0/4\lambda\sim 1$, in~(20) the second term 
dominates the first one and we obtain a small AHP confined to
a single site. Contrary, for $\lambda=1.5$ and $\alpha=3$,
we are still in the large polaron region~II due to the 
high phonon frequency (cf. Fig. ~1 and the spatial extension of the 
static lattice distortion shown in the left inset of Fig.~3~(b)). 
Nevertheless the correlations between the electron and 
the phonon remain local. But now,   
since the $\mD_j/4\lambda$ are small for all $j$,  
the peak structure of  $\chi_{0,j}$ at $j=0$ results 
from the first term in~(20). That means it is mainly 
triggered by the $\gamma$--effect ($\gamma\simeq 0.69$).
This interpretation is substantiated by our ED results yielding 
a mean phonon number in the ground state of about 0.625, which  
clearly indicates that the zero--phonon state is still the most 
probable one. Therefore the approximation we applied 
by deriving~(10) and~(20) is justified.
\subsection{Optical response}
Extremely valuable informations on the low--energy excitations 
in interacting electron/exciton--phonon systems can be obtained 
by studying their optical response.   
Actually, the optical absorption of small polarons is distinguished
from that of large (or quasi--free) polarons by the shape and the 
temperature dependence of the absorption bands which arise from exciting
the ST carrier from or within the potential well that binds it~\cite{Emi95}.  
Furthermore, as was the case with the ground state properties, 
the optical spectra of light and heavy electrons/excitons
differ essentially as well~\cite{Ra82}. 
In the most simple weak--coupling and non--adiabatic strong--coupling limits,
the absorption associated with photoionization of Holstein polarons is well
understood and the optical conductivity can be analyzed 
analytically~\cite{Emi93,Mah90,RH67,Lo88,Feea94} 
(for a detailed discussion of small polaron transport phenomena 
we refer to the review of Firsov~\cite{Fi95}).   
Serious problems, caused, for example, by the complicated behaviour
of the adiabatic potential surface, arise if one tries to calculate   
the spectrum of self--trapped light excitons~\cite{Ra82}. 
Moreover, the intermediate coupling and frequency regime 
is as yet practically inaccessible for a rigorous analysis.
On the other hand, previous numerical studies of the optical 
absorption in the Holstein model were limited to very 
small 2-- or 4--site clusters~\cite{AKR94,CSG97}.   
In order to calculate the optical conductivity numerically 
in a wide parameter range on fairly large systems,  
we have implemented our computer code, 
which is based on a combination of Lanczos diagonalization, 
Chebyshev recursion and maximum entropy methods~\cite{Siea96}, 
on parallel machines.  

The real part of the optical conductivity, 
$\mbox{Re}\sigma(\omega)={\cal D}\delta(\omega)+\sigma^{reg}(\omega)$, 
can be decomposed into the Drude weight $\delta$--function at $\omega=0$
and a regular part ($\omega>0$) written in 
spectral representation at $T=0$ as~\cite{Da94}
\begin{equation} 
\sigma^{reg}(\omega)=\sum_{m\neq 0}
\frac{|\langle {\mit \Psi}_0^{} |i \sum_{j}( a_{j}^{\dagger}
 a_{j+x}^{} - a_{j+x}^{\dagger}a_{j}^{}) |  {\mit \Psi}_m^{} 
       \rangle |^2}{E_m-E_0} \;\delta(\omega -E_m+E_0)\,.
\end{equation}
In~(21), $\sigma^{reg}(\omega)$ is given in units of $\pi e^2$ and we have
omitted the $1/N$ prefactor. For the discussion of the
optical properties it is useful to consider the spectral weight function  
\begin{equation}
\cS^{reg}(\omega)=\int_0^{\omega}d\omega^{\prime}\sigma^{reg}
(\omega^{\prime})
\end{equation} 
as well.

Numerical results for both, $\sigma^{reg}(\omega)$ and $\cS^{reg}(\omega)$,
are presented in Fig.~4. We will start with the somewhat more
subtle case of {\it light electrons}. 

But first let us recall that, 
restricting ourselves to phononic states $|s\rangle_{ph}=\prod_{i=0}^{N-1}
\frac{1}{\sqrt{n_i^s!}}\left(b_i^\dagger\right)^{n_{i}^{s}}\,|0\rangle_{ph}$
with at most $M$ phonons, a $\bk$--symmetrized state of the Holstein model is 
given as $|{\mit \Psi}_{\bk}^{}\rangle = \sum_{m=0}^M
\sum_{\bar{s}=1}^{\bar{S}(m)} c_{\bk}^{m,\bar{s}} \,
|K;m,\bar{s} \rangle$, where $\bar{S}(m)=(N-1+m)!/(N-1)!m!$
(for more details see~\cite{FLW97}). $\bk$ denotes the {\it total} momentum
of the coupled EP system. Then, if the EP coupling is finite,
the ground state $| {\mit \Psi}_0^{}\rangle$ and all excited states  
$| {\mit \Psi}_m^{}\rangle$ contain components that correspond to 
$m$--phonon states (with $m=\sum_{i=0}^{N-1} n_i^s \le M$,  $n_i^s\in [0,m]$) 
in the tensorial product Hilbert space of electronic and phononic states. 
When the EP coupling is small ($\lambda\ll 1$), these multi--phonon states 
have less spectral weight, i.e., the phonon distribution of the ground state,  
$|c^m_0|^2(M)=\sum_{\bar{s}}  |c_{\bk=0}^{m,\bar{s}}|^2$ ,  
exhibits a pronounced maximum at the zero--phonon state~\cite{FLW97}.
The maximum of $|c^m_0|^2(M)$ is shifted to larger values of $m$ as
$\lambda$ increases.  

Keeping this in mind and notice further that in~(21) an optical transition can 
take place only within the $\bk=0$ sector ($|{\mit \Psi}_0^{}\rangle\equiv
| {\mit \Psi}_{\bk=0}^{}\rangle$), 
the peak structure of $\sigma^{reg}$ shown in 
Fig~4~(a) may be easily understood in connection with the single particle 
spectrum. For low phonon frequencies ($\alpha\ll W$), the energy to excite 
one phonon lies inside the bare tight--binding band $E_K^{(0)}=-2t \cos K$ 
and we observe a flattening of the coherent band structure  $E_K$
at large momenta~\cite{WF97,St96}. Then the coherent bandwidth 
$\mD E=E_{\pi}-E_0$ is approximately given by $\alpha$ 
($\ll [E_K^{(0)}-E_0^{(0)}]$), 
i.e. by the phonon frequency, where the 
states at finite momenta are predominantly ``phononic'' states with 
less ``electronic'' spectral weight. Thus, although in principle an 
optical excitation can be achieved by ``adding'' phonons with 
opposite momentum to these states (in order to reach the $\bk =0$ sector),
the overlap to the mainly ``electronic'' ground state is extremely small.
Therefore we found, roughly speaking, the first transitions 
with non--negligible weight to the free electron states and 
its vibrational satellites (see $S^{reg}(\omega)$). 
This is perfectly illustrated by the inset of Fig.~4~(a).
Here the first and second group of peaks is approximately 
located at the bare tight--binding 
energies, $E^{(0)}_{\bk}$~(+ $n\cdot\alpha$),  
for the allowed wave vectors of the 8--site lattice used 
in the numerical calculation ($\bk =\pi/4$,  $\pi/2,\ldots$).
Of course, with increasing the lattice size the $K$ values will become dense 
and we will obtain the monotonous decay of the optical absorption coefficient
observed for large polarons above the photoionization threshold~\cite{Emi95}.  
To understand the changes in the optical absorption  
in the crossover region from a large--size polaron to small AHP, 
the main part of Fig.~4~(a) shows $\sigma^{reg}$ 
at two intermediate EP coupling parameters. 
In this case the coherent band structure $E_K$ gets stronger renormalized, 
but, more important, the phonon distribution function in the ground state, 
$|c^m_0|^2(M)$, becomes considerably broadened.  
For instance, at $\alpha=0.1$, we have   
$|c^m_0|^2(M=25)\simeq 0.008$ ($m=0$), 0.095 ($m=7$), 0.008  ($m=18$)
and $|c^m_0|^2(M=25)\simeq  0.0002$  ($m=0$), 0.1 ($m=12$), 0.0005  ($m=24$)
for $\lambda=0.9$ and  $\lambda=1.0$, respectively.  
Therefore the overlap with excited multi--phonon states is enlarged
and the optical response is enhanced.  
The lineshape of the absorption bands reflects the phonon distribution
in the ground state, where the small oscillations are due to the 
discreteness of the phonon frequency. 
As a result, the peak structure is smeared out and 
the wide side--bands,  belonging to different electronic momenta 
(e.g., $K=\pi/4$ and  $K=\pi/2$), merge with each other. 

We now turn to the {\it heavy electron} case (Fig.~4~(b)). 
The inset again illustrates the behaviour in the large polaron regime 
($\lambda=1,\;\alpha=3$; cf. Fig.~1), where 
a rather moderate band renormalization occurs ($\mD E \sim 2.33$
due to the flattening effect~\cite{WF97}). 
But now the phonon frequency is large compared to the 
finite--size gaps between the first mainly ``electronic'' excitations  
($E_K-E_0 < \alpha$ for $K\leq\pi/2$). Therefore, in contrast
to the light electron case (cf. Fig.~4~(a)), the different 
absorption bands (each built up by several ``electronic'' $K$-levels) 
can be classified according to the number of phonons involved in the 
optical transition. 
As can be seen from the main part of Fig.~4~(b), the 
absorption spectrum for a small--size polaron 
is quite different from that of a large polaron. 
According to the results of Sec.~III~A, a small NLFP 
is formed in the strong--coupling non--adiabatic limit. 
Here the phonons will heavily dress the electron and, concomitantly, 
the ``electronic character'' of the resulting 
strongly renormalized small polaron band 
becomes suppressed (cf. the discussion of the  
$K$--dependent wave--function renormalization factor  
${\cal Z}^{(a)}_K=|\langle {\mit \Psi}_K^{}|a_K^{\dagger}| {\mit \Psi}_0^{}
\rangle|^2$ in Ref.~\cite{FLW97}).
For our parameters ($\lambda =6$, $g^2=4$),
the maximum in phonon distribution function is located 
between $m=3$ and 4. The renormalized bandwidth 
is small compared to all other energy scales 
($\mD E\sim 0.0782\ll\alpha,\,W$). 
Since the current operator connects only states having 
substantial overlap as far as the phononic state is concerned, 
multi--phonon absorptions (i.e., non--diagonal transitions~\cite{Mah90}) 
become now increasingly important in the optical response. 
This leads to the peak structure observed
for the non--adiabatic small polaron optical conductivity in Fig.~4~(b).
Obviously, the different bands are being separated by multiples of the bare 
phonon frequency. The height of the jumps in 
the $\omega$--integrated conductivity is directly related 
to the probability of the corresponding absorption process. 
We found that substantial spectral weight stays in the 
lower energy part of the spectrum at frequencies comparable
to $\ep=2\lambda\; (\simeq  m_{max}\cdot\alpha)$.
These absorptions, resembling to some extend a 
large polaron's absorption, are signatures 
of a ST polaron with intermediate size. 
In the extreme strong--coupling limit the dominant absorption process
results from the transfer of the ST carrier to the neighbouring site 
without changing the lattice distortion. That means, the optical absorption 
spectrum should exhibit a single--peak structure at $\omega=2\ep=4\lambda$, 
which corresponds to the lowering of the electronic energy
associated with the small--polaron formation~\cite{Kl63,RH67}.
This feature already evolves for the coupling strength
considered in Fig.~4~(b) (cf. $\mD S^{reg}$ for 
the $7^{th}$ absorption band).  
\subsection{Kinetic energy} 
Further information about the transition from large to small polarons 
can be obtained from the behaviour of the polaron kinetic energy $E_{kin}$. 
Replacing $\bar{\cH}$ by $\cH$, the kinetic energy  can be easily obtained 
from static correlation function~(15).  
On the other hand, according to the f--sum rule, 
$E_{kin}$ is directly related to the $\omega$--integrated 
optical conductivity, 
\begin{equation}
-\frac{E_{kin}}{2} =\cS^{tot}= \frac{{\cal D}}{2\pi e^2} + \cS^{reg}\,. 
\end{equation}
Calculating,  via (21) and~(22), $\cS^{reg}=\cS^{reg}(\infty)$ numerically,
allows us to determine the Drude weight ${\cal D}$ as well. 
Sometimes one defines an effective polaronic transfer 
amplitude~\cite{FRWM95,MR97}, $t_{p,eff}=E_{kin}(\lambda)/E_{kin}(0)$,
in order to characterize the polaron mobility. In our reduced units 
we have $t_{p,eff}\equiv\cS^{tot}$. From~(23) it is obvious that
$t_{p,eff}$ includes both coherent and incoherent transport processes.
Hence $t_{p,eff}$ substantially differs from the exponential 
factor $\mbox{e}^{-g^2}$, obtained in lowest order perturbation theory, 
and cannot be used to determine the coherent bandwidth $\mD E$ under
strong coupling conditions ($\lambda,\,g^2\gg1$). 

The evolution of the kinetic energy ($\cS^{tot}$)
as a function of the EP coupling $\lambda$ is displayed 
in Fig.~5~(a) and~(b) for the case of light and heavy electrons, 
respectively.  In agreement with previous numerical 
results~\cite{RL82,FRWM95,MR97,JW97},  
$E_{kin}$ clearly shows the crossover from a large polaron, 
characterized by a $\cS^{tot}$ that is only weakly reduced from 
its non--interacting value ($\cS^{tot}(\lambda =0)=1$), to a less 
mobile small AHP/NLFP in the adiabatic/non--adiabatic strong--coupling limit.
 
For {\it low phonon frequencies} ($\alpha=0.1$; Fig.~5~(a)), 
we found a rather narrow transition region. 
As recently pointed out by Capone~et~al.~\cite{CSG97},
the decrease of $\cS^{tot}$ in the crossover regime
is driven by the sharp fall of the Drude weight. 
By contrast the optical absorption due to
inelastic scattering processes,  described by the regular part of 
the optical conductivity, becomes strongly enhanced~\cite{FLW97} 
(see the behaviour of $\cS^{reg}$). 
It is worth emphasizing that the IVLF--results for $\cS^{tot}(\lambda)$ 
are in excellent agreement with the ED and DMRG data. 

The large to small polaron transition is considerably broadened
at {\it high phonon frequencies} ($\alpha=3.0$; Fig.~5~(b)). 
Here the IVLF--results start to deviate from the exact ones
when $g^2$ gets much larger than one, thus making the
lowest--order zero--phonon approximation inherent in the IVLF--scheme 
less justified. As mentioned already in Sec.~III~A, the
non-analytic jump-like behaviour at $\lambda\sim 4$ is an apparent shortcoming
of the variational approach  which compares ground--state energies only.   

Although for large enough $g$ and $\lambda$ the simple formula 
$\bar{W}=W\exp\{-g^2\}$,
which should not be identified with 
the ``Lang--Firsov approach''~\cite{MR97,FK97}, 
works perfectly well in the determination of the {\it coherent
bandwidth} ($\mD E\simeq\bar{W}$)~\cite{AKR94,FLW97},  
the need of going beyond  the lowest order of approximation to
obtain reliable results for the {\it kinetic energy} has been emphasized many
times~\cite{Fi75,FLW97,JW97,AM95,FK97}. In the non--adiabatic
strong--coupling limit ($g^2\gg 1$, $\alpha>1$),
the ground--state energy obtained within second--order
perturbation theory is a tiny little bit
lower than the IVLF-energy and almost coincides with the ED result.
Adapting the second--order strong--coupling approach  
presented in our previous
 work~\cite{FLW97} to the 1D case (with $\gamma=1$),  
the kinetic energy is obtained, via 
$E_{kin}^{SCPT}=t\,\partial_t \langle \cH \rangle$, 
as     
\begin{equation}
E_{kin}^{SCPT}=
-\frac{4}{\alpha} \Big\langle\frac{1}{s}\Big\rangle_{\kappa=2g^2}^{}
-\mbox{e}^{-g^2}\left[ 2 
+ \frac{4}{\alpha}\Big\langle\frac{1}{s}
\Big\rangle_{\kappa=g^2}^{}\right]\,.
\end{equation}
Here $\langle \ldots\rangle_\kappa$ means the average with respect
to the Poisson distribution with parameter $\kappa$.
As can be seen from Fig.~5, at large EP interactions,  
the strong--coupling perturbation theory
(SCPT) gives a sufficiently accurate description of $\cS^{tot}$
in both the adiabatic and non--adiabatic regimes.
\section{Summary}
The main objective of this study was to re--examine
in detail the self--trapping transition of electrons and excitons 
in one dimension (1D) within the framework of the Holstein model 
by the use of computational techniques. The calculations are
performed by exact diagonalizations of finite systems, where the 
full dynamics and quantum nature of phonons was taken into account. 
Therefore the results are unbiased and allow to test the applicability 
of a numerically much more efficient variational (IVLF) Lanczos approach 
proposed by the authors. This IVLF--Lanczos technique is designed 
to analyze the ground--state properties of strongly correlated 
electron--phonon (EP) models on fairly large lattices, including 
static displacement field, non--adiabatic polaron and squeezing effects. 

Let us summarize the main outcome of our work.

Fig.~6 illustrates the basic physics 
contained in the single--particle Holstein model. 
This model describes an continuous transition
from large to small polarons with increasing EP coupling strength.
Depending on the adiabaticity of the system, $\alpha=\omega_0/t$, 
the crossover regime is determined by the more stringent of the
two conditions $\lambda=\ep/2Dt$ and $g^2=\ep/\omega_0$. 
Thus starting from  ``light'' ($\alpha<1$) or ``heavy''  ($\alpha>1$)
electrons  it is possible to understand the formation of small 
adiabatic ``Holstein'' (AHP) or non--adiabatic ``Lang--Firsov'' (NLFP)
polarons as two limiting cases of a general picture. 

In the infinite 1D Holstein model the self--trapped  state 
is the ground--state for any value of the EP coupling. 
By contrast, in a finite 1D Holstein system a critical 
length or equivalently a critical EP coupling strength 
exists for self--trapping, rigorously at least 
at $\omega_0=0$. This may be of importance for 
spatially restricted systems like $\rm C_{60}$.
The large--size polaron is characterized by spatially extended 
lattice deformations. Within our IVLF--description, the variation 
of the displacement fields follows the (adiabatic) formula 
$\mD_i\propto \mbox{sech}^2 [\lambda_{eff} \cdot i]$  even in the 
non--adiabatic regime, but with a strongly reduced inverse length scale
given by an effective coupling constant 
$\lambda_{eff}(\lambda,\alpha)\ll\lambda $. 
The mean phonon number in the large polaron ground state is rather small.  
In 2D and 3D Holstein systems a large polaron phase does not exist.  

The small polaron state is basically a multi--phonon state characterized
by strong on--site electron--phonon correlations. Due to a large 
local static lattice distortion the AHP becomes quasi--localized on a 
single site. Also the NLFP is immobile because it has to drag with it 
a large number of phonons in its phonon cloud.     

The transition from large to small polarons is accompanied 
by significant changes in their optical response. Especially for 
``light'' electrons the spectral weight of the regular part of 
the optical conductivity is strongly enhanced at the transition.
For ``heavy'' electrons the lineshape of the optical absorption 
spectra is highly asymmetric in the intermediate--to--strong coupling 
regime and therefore differs considerably from the usual  
small polaron hopping behaviour obtained for $g^2\gg 1$.  

As a result of self--trapping the mobility of the charge carriers is reduced.  
The kinetic energy indicates that the crossover region from
large to small ST states is rather narrow (broad) in the  
adiabatic (non--adiabatic) regimes. 
The formation of small adiabatic Holstein polarons is accompanied by 
a dramatic kinetic energy loss mainly driven by a sharp drop 
of the Drude weight. Since both the bandwidth and the 
(electronic) spectral weight of the small polaron band 
are exponentially reduced  in the extreme strong coupling
case  $g^2,\,\lambda\gg1$, coherent small polaron transport 
becomes rapidly destroyed by thermal fluctuations.

Finally, we would like to stress that the simple IVLF--Lanczos approach  
provides an exceptional good description of the ground--state 
properties of the Holstein model, in particular for the light electron case.
\acknowledgments
The authors would like to thank J. Loos and D. Ihle
for many helpful discussions.
We are greatly indebted to E. Jeckelmann and S. R. White for putting their
DMRG data at our disposal.  
Special thanks go to the LRZ M\"unchen, HLRZ M\"unchen, and the HLR 
Stuttgart for the generous granting of their computer facilities.
This work was performed  under the auspices of Deutsche 
Forschungsgemeinschaft, SFB 279.
\narrowtext\def\baselinestretch{0.95}
\bibliography{ref}
\bibliographystyle{phys}

\figure{FIG. 1. Phase diagram of the 1D Holstein model. Nearly free
polarons, large polarons, and small polarons exist in the regions
I, II, and III, respectively. Results are obtained for finite rings with 
$N=32$ $(\circ )$, 64  $(\Box,\;\ast  )$, 96 $(\bigtriangleup )$,
and 128 $(\bigtriangledown)$ sites using the IVLF--Lanczos method.
The inset shows the critical coupling for self--trapping, $\lambda_c$,
at $\alpha=0$ $(\times )$, 0.1 $(\Diamond )$, 1.0  $(\triangleright )$, 
and 3.0   $(\triangleleft )$. The solid line gives the relation~(17).
For further explanation see text.} 
\figure{FIG. 2. Variation of the displacement fields $\mD_i$ away 
from the central site $0$ for several characteristic EP couplings $\lambda$
in the adiabatic (a) and non--adiabatic (b) regimes. The dependence of 
the Lang--Firsov polaron variational parameter 
$\gamma$ on $\lambda$ is depicted in (c).}
\figure{FIG. 3. Electron--lattice correlations $\chi_{0,j}$ in the
adiabatic weak--to--intermediate EP coupling (a) and 
non--adiabatic intermediate--to--strong EP coupling (b) cases.
IVLF--Lanczos results are compared with ED data and the DMRG results
taken from Ref.~\protect\cite{JW97}.} 
\figure{FIG. 4. Optical absorption in the 1D Holstein model. 
For $N=8$ and $M=25$, the regular
part of the conductivity $\sigma^{reg}$ (thin lines) and the 
integrated spectral weight $S^{reg}$ (thick lines) are plotted as a function of
$\omega$ at different EP couplings for the ``light'' (a) and ``heavy'' (b) 
electron cases.}  
\figure{FIG. 5. Kinetic energy [in units of $(-W)$], $S^{tot}$, and
contribution of $\sigma^{reg}$ to the f--sum rule,  $S^{reg}$, as a 
function of the EP coupling, $\lambda$, in the adiabatic (a) and  
non--adiabatic (b) regimes. Circles (stars) denote exact 
(DMRG~\protect\cite{JW97}) data obtained for a lattice 
with $N=8$ (32) sites. IVLF--Lanczos results (solid curves) 
are compared with the predictions of standard first--order 
($\bar{W}/W=\exp\{-g^2\}$; thin solid line) and  second order 
(chain--dashed curves) perturbation theory.}
\figure{FIG. 6 Schematic phase diagram of the Holstein model.}
  \end{document}